\documentclass[%
 aip,
 amsmath,amssymb,
 reprint,%
twocolumn,
floatfix,
]{revtex4-1}

\usepackage{graphicx}
\usepackage{amssymb}
\usepackage{amsmath}
\usepackage{latexsym}
\usepackage{bbold}
\usepackage{bm}
\usepackage{color}
\usepackage{soul}

\begin{document}
\title{Neoclassical transport of tungsten ion bundles in total-f neoclassical gyrokinetic simulations of a whole-volume JET-like plasma}

\author{J. Dominski}
\email{jdominsk@pppl.gov}
\affiliation{Princeton Plasma Physics Laboratory, 100 Stellarator rd, Princeton 08540 New Jersey, USA}

\author{C.S. Chang}
\affiliation{Princeton Plasma Physics Laboratory, 100 Stellarator rd, Princeton 08540 New Jersey, USA}
\author{R. Hager}
\affiliation{Princeton Plasma Physics Laboratory, 100 Stellarator rd, Princeton 08540 New Jersey, USA}
\author{S. Ku}
\affiliation{Princeton Plasma Physics Laboratory, 100 Stellarator rd, Princeton 08540 New Jersey, USA}
\author{E.S. Yoon}
\affiliation{Ulsan National Institute of Science and Technology, Ulsan 44919, South Korea}
\author{V. Parail}
\affiliation{Culham Centre for Fusion Energy, Culham Science Centre Abingdon, OX14 3DB, United Kingdom}

\begin{abstract}
The application of a bundling technique to model the diverse charge states of tungsten impurity species in total-f gyrokinetic simulations is demonstrated. The gyrokinetic bundling method strategically groups tungsten ions of similar charge, optimizing computational efficiency. The initial radial configuration of these bundles and their respective charges are derived from a coronal approximation and the quasi-neutrality of the plasma. A low-density JET H-mode like plasma is simulated using the neoclassical version of XGC across the entire plasma volume, spanning from the magnetic axis to the divertor. An accumulation of tungsten is observed at the pedestal top, as a result of low-Z tungsten ions moving inward from the scrape-off-layer (SOL) into the core region and high-Z tungsten ions moving outward from the core into the pedestal. This organization of the fluxes cannot be captured by a single tungsten-ion simulation. Large up-down poloidal asymmetries of tungsten form in the pedestal and strongly influence the direction of neoclassical fluxes. The temperature screening effect and its correlation with asymmetries is analyzed. 
\end{abstract}

\maketitle

\section{Introduction}
Understanding the neoclassical and turbulent impurity transport and its impact on the main plasma confinement is an important subject in magnetic fusion research. Seeding of impurity particles was found to improve the plasma confinement in the so-called ``radiation improved mode'' (RI-mode) of operation~\cite{Weynants94} and, {more recently,} tungsten (W) impurities have been found to degrade the pedestal confinement of JET H-mode plasma~\cite{Litaudon17}. Whole-volume modeling of impurities transport with the X-point gyrokinetic code XGC~\cite{Chang17} could bring new insights into this topic.

A previous XGC study~\cite{Kim17} showed that the presence of carbon (C) impurities could improve the main-ion confinement of tokamak plasma by reducing the ITG turbulence amplitude. In addition to the usual main-ion dilution effect, this confinement enhancement was caused by both the impurity-wave interaction and the enhancement of the mean $E\times B$ shearing rate. Finally, the central peaking of carbon measured in turbulence simulations was milder than the one predicted by neoclassical theories.

The importance of simultaneously considering neoclassical and turbulence physics has been studied in Ref.~\cite{Esteve18} with the GYSELA code. The nonlinear interplay between neoclassical and turbulence dynamics underscores the necessity for a consistent modeling approach to precisely predict transport properties. In these simulations, the electrons were assumed adiabatic and the magnetic geometry had no shaping or X-point.

Reduced tools that simplify the problem by treating neoclassical and turbulence physics separately~\cite{Angioni14,Angioni15,Loarte15} have been, nonetheless, meaningful for predicting different features of transport or for confirming earlier theories~\cite{Chang83,Hinton85,Helander98,Romanelli98}: the formation of tungsten asymmetries due to strong centrifugal forces has been studied in Ref.~\cite{Angioni14}, the impact of poloidal asymmetries on tungsten transport has been studied in the core of a JET-model plasma in Ref.~\cite{Angioni15}, the collisional up-down asymmetry of the electrostatic potential is discussed in Refs.~\cite{Hinton73,Wong09}, and more recently the influences of the rotation and electrostatic potential asymmetry on the neoclassical tungsten flux is discussed in Refs.~\cite{Maget20a,Maget20b,Maget22}.

In our work, we present gyrokinetic neoclassical tungsten impurity transport properties (or collisional transport modeled with the gyrokinetic ordering) in a low-density model JET H-mode plasma, optimized for the study of the temperature-gradient screening effect. Tungsten ions are modeled with several bundles, covering the whole plasma volume with the current version of XGC that includes impurities~\cite{Dominski19a}. Full-f collisions and electrostatic interactions between drift-kinetic electrons, gyrokinetic ions, and all gyrokinetic tungsten bundles are fully accounted for. Unlike bundling techniques in fluid codes~\cite{Strachan2011}, our gyrokinetic bundling involves a fixed electric charge for each gyrokinetic bundle. The profile of averaged electric charge of tungsten ions is thus modeled with a few gyrokinetic bundles, evolving in time according to the total-f gyrokinetic equation. 

Our gyrokinetic simulations, incorporating multiple ionization states of tungsten organized in bundles, emphasize the significance of modeling these diverse charge states. Low-Z and high-Z tungsten ions exhibit distinct dynamics, contributing to a complex organization of the tungsten flux that cannot be adequately captured by a single tungsten ion species. In these simulations, the accumulation of tungsten at the pedestal-top results from a combination of tungsten penetration from the scrape-off layer (SOL) and an outward flux from the core, both influenced by temperature screening effects. It is noteworthy that, although our model JET plasma differs from experimental discharge, an accumulation of tungsten in the pedestal of JET is observed in Ref.~\cite{Kochl18}.

The paper is organized as follows.  In Sec.~\ref{sec:model}, we provide a detailed description of the impurity model employed in XGC. Sec.~\ref{sec:JETcase} introduces the JET-like test case. Results of neoclassical XGC simulations with a single ionized tungsten species are presented in Sec.~\ref{sec:singleW_results}, and with multiple ionized tungsten species in Sec.~\ref{sec:multiW_results}. The influence of tungsten on the neoclassical main ion transport is reported in Sec.~\ref{sec:ionfluxes}. Finally, a conclusion is drawn in Sec.~\ref{sec:conclusion} together with a discussion on future work.

\section{Model used in the multi-species version of XGC}
\label{sec:model}
\subsection{Brief description of XGC}
XGC is a total-f gyrokinetic particle in cell (PIC) code that simulates the plasma in the whole tokamak volume including the scrape-off-layer and divertor regions~\cite{Ku16,Ku18,Hager2022}.

In the current manuscript we use the axisymmetric version of XGC, often referred to as XGCa. This axisymmetric version of XGC does not include the turbulence physics and solves for an axisymmetric gyrokinetic Poisson equation of the form
\begin{equation}
    \mathcal{L} \phi^{(n=0)}=\left(-n_e+Z_a\sum_{a}\bar{n}_a\right)^{(n=0)},
\end{equation}
where the superscript $^{(n=0)}$ indicates that we consider the toroidaly axisymmetric quantity, and where $a$ labels each gyrokinetic ion. Details concerning the gyrokinetic polarization operator $\mathcal{L}$ are provided in Ref.~\cite{Dominski19a}.

XGCa, which uses a 5-dimensional full-f Fokker-Planck collision operator~\cite{Eisung14,Hager16}, is thus an appropriate tool for studying neoclassical physics in the whole device. Compared to kinetic neoclassical codes that are based on the drift-kinetic ordering, XGCa uses the gyrokinetic ordering that includes finite gyroradius effects. As pointed out in Ref.~\cite{Vernay10} more terms are accounted for than in the usual neoclassical drift ordering. This makes gyrokinetic codes a powerful tool to study collisional physics in regions of steep gradient, near the magnetic axis, near the X-point, or anywhere where the drift ordering can be problematic. The guiding-center equations of motion, here expressed in the $v_\|$ formalism, read
\begin{equation}
\begin{cases}
\dot{\mathbf{X}}=\frac{1}{B_{\parallel}^\star}\left(v_\parallel \bm{B}^\star-\mu\nabla B\times \bm{b}+
q\bar{E}^0\times\bm{b}\right)\\
\dot v_\parallel =-\frac{\dot{\mathbf{X}}}{mv_\|}\cdot(\mu\nabla B+q\bar{E}^0)
\end{cases},
\label{eq:GKE}
\end{equation}
where $\bar{E}^0=-\nabla \langle{\phi^{(n=0)}}\rangle_\alpha$ is the axisymmetric electric field and $\langle \ \ \rangle_\alpha$ is the gyroaveraging operation. 

In XGCa, all the ion species are modeled with a complete gyrokinetic species, whose marker particles are pushed with all drift terms and are subject to gyroaveraging. The electrons are drift-kinetic which means they are also pushed with all drift terms but their Larmor radius is zero, \textit{i.e.},  $\langle\phi\rangle_\alpha=\phi$. This multi-ion species version of the code is described in Ref.~\cite{Dominski19a}. 

Now that we clarified the gyrokinetic dynamics of the marker particles, let us point that we evolve the particle distribution function $f$ with a gyrokinetic equation of the form
\begin{equation}
\frac{\partial f}{\partial t}+\dot{\mathbf{X}}\cdot\frac{\partial f}{\partial {\mathbf{X}}}+\dot{v}_\parallel\frac{\partial f}{\partial v_\parallel}=\mathcal{C}[f],
\label{eq:GKE_n0}
\end{equation}
where the guiding-center Coulomb collisions are modeled with a multi-species nonlinear Fokker-Plank operator, $\mathcal{C}[f]=\sum_{a,a^\prime}\mathcal{C}[f_a,f_{a^\prime}]$, as described in Refs.~\cite{Eisung14,Hager16}. Note that no source term (heat, torque, etc...) has been used in this manuscript. Their influence will be studied in a future work.

Let us emphasize that all (axisymmetric) interactions between, ions, electrons, and bundles are accounted for. This includes the full-f collisions between all species (electrons, ions, and bundles) as well as the interaction through the electrostatic field (GAM, poloidal field asymmetries,...). Indeed, each ion species and each tungsten bundle species is modeled with a gyrokinetic species. 
\begin{itemize}
    \item 
    For example, in a plasma that includes electron, deuterium, and 4 bundles of tungsten, each one of these 6 kinetic species collides with itself and with the 5 others ($\mathcal{C}=\sum_{a,a^\prime}\mathcal{C}[f_a,f_{a^\prime}]$ with $a$ and $a^\prime$ any kinetic species).
\item Moreover, each gyrokinetic species contributes to the density perturbation in the right-hand-side of the gyrokinetic Poisson equation. This includes the bundles, such that any small or large 2D poloidal asymmetries developing on a bundle will self-consistently affect the electrostatic field asymmetry through the gyrokinetic Poisson equation and interact with the motion of all other marker particle species. 
\end{itemize}

Finally, let us briefly close the description of our gyrokinetic model by mentioning that in XGC the particle distribution function is modeled with a total-f representation of the form
\[
f(t):=f_0(t)+\widetilde{\delta f},
\]
where $f_0(t)$ is a slowly relaxing background and $\widetilde{\delta f}$ is a Particle-in-Cell (PIC) representation of the perturbation $\delta f$. In comparison, the usual delta-f PIC representation uses a fixed background $f_0$.

The perturbation $\widetilde{\delta f}$ is represented by marker particles that carry the information of the initial background $f_0$ and of the perturbation $\delta f$ with two weights:
\begin{eqnarray}
w_0 &:=& f_0(t=0)/g,\\
w_1 &:=& \delta f / (w_0 g),
\end{eqnarray}
where $g$ is the distribution of marker particles. The weight $w_0$ is fixed and the weight $w_1$ evolves according to the different terms on the right-hand-side of the gyrokinetic equation
\begin{equation}
\frac{\partial \delta f}{\partial t}+\dot{\mathbf{X}}\cdot\frac{\partial \delta  f}{\partial {\mathbf{X}}}+\dot{v}_\parallel\frac{\partial \delta  f}{\partial v_\parallel}=-\frac{df_0}{dt}+\mathcal{C}[f].
\label{eq:GKE_df}
\end{equation}
In XGC, the implementation of the background term $\frac{-df_0}{dt}$ employs an additional weight, $w_2$, to compute the collision-less evolution
\[
\dot w_2 =-\frac{1}{w_0g}\frac{df_0}{dt},
\]
with the direct delta-f weight evolution scheme
\begin{equation}
w_2^{t+\Delta t}=w_2^{t} +\frac{1}{w_0 g}\frac{f_{0p}^t-f_{0p}^{t+\Delta t}}{\Delta t},
\label{eq:weightevolution}
\end{equation}
where $\Delta t$ is a time step and $f^{t}_{0}=f_0(\bm{X}^{t},v_{\|}^{t},\mu^{t};t)$ is the value of the background distribution function at the position of the  considered particle at time $t$. Contributions from the collision term are computed in a second step after upgrading the weight $w_1$, with 
\[
w_1^{t+\Delta t}=w_1^{t}+(w_2^{t+\Delta t}-w_2^{t})\underbrace{+\frac{\mathcal{C}}{w_0 g}}_{\text{computed after}}.
\]
  
The total-f model enables the relaxation of the background distribution function $f_0(t)$ while conserving the total distribution function $f(t)$. This relaxation of $f_0$ is applied on a slow time scale by transferring part of the axisymmetric relaxation contained from $\widetilde{\delta f}$ into $f_0$ with an operation of the form
\[
f_0:= f_0 + \alpha\, \left( \widetilde{\delta f}\right)^{(n=0)}
\]
and
\[
\widetilde{\delta f}:= \widetilde{\delta f} - \alpha\, \left( \widetilde{\delta f}\right)^{(n=0)},
\]
where $\alpha$ is a small input parameter, see Ref.~\cite{Ku16}. Among other things, this evolution of the background enables to reduce the particle noise present on $\widetilde{\delta f}$ and it provides a more consistent modeling of the weight evolution equation, Eq.~\eqref{eq:weightevolution}, than if using a fixed $f_0$ background.

In the simulations presented in this manuscript, the relaxation of the background is modeled by a 4D grid representation $f_g=f_0(t)-f_0(t=0)$ and 
\[
f(t)=\underbrace{f_0(t=0)+f_g}_{f_0(t)}+\widetilde{\delta f}.
\]

\subsection{Grouping of multiple ionization states in ``bundles''}
\label{sec:multiple_ionization}

Tungsten impurity ions potentially exist in $74$ ionization states ($Z=1,...,74$) in a hot plasma. Simulating so many electric charge states as individual species is out of reach for recent super-computers. Instead, we model these ionization states with several gyrokinetic bundles (a minimum of $4$ in this paper). Each bundle is modeled as an independent gyrokinetic ion species. Compared to bundled impurities in fluid models, this gyrokinetic approach resolves the  kinetic dynamics which is not captured by fluid codes. Indeed, each bundle is evolved according to the gyrokinetic XGC model as if it was a single species. Atomic processes (ionization/recombination) between bundles are not included in the present work and will be considered in a future work.

\begin{figure}
      \includegraphics[width=8.5cm]{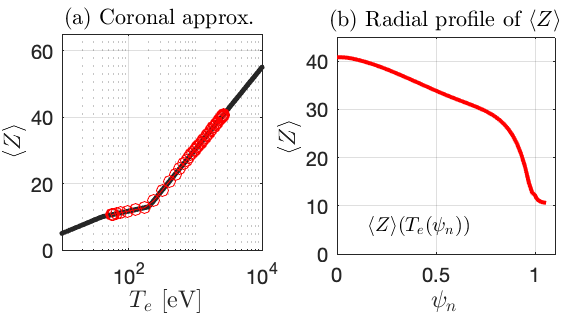}
      \caption{Average ionization of tungsten given in the Coronal approximation. (a) Coronal approximation of the average charge $\langle Z\rangle$ as a function of the electron temperature, $T_e$. This curve is approximated from Ref.~\cite{Demura15}. (b) Profile of average charge $\langle Z\rangle$, by using the data in (a) and the temperature profile $T_e(\psi_n)$ from Fig.~\ref{fig:JET_eD_vs_eDW_equilibirum_poster}(c). The red circle markers in (a) represent the points used for mapping the curve in (b).  }
      \label{fig:coronal_approximation}
      \end{figure}

\begin{figure}
      \includegraphics[width=8.5cm]{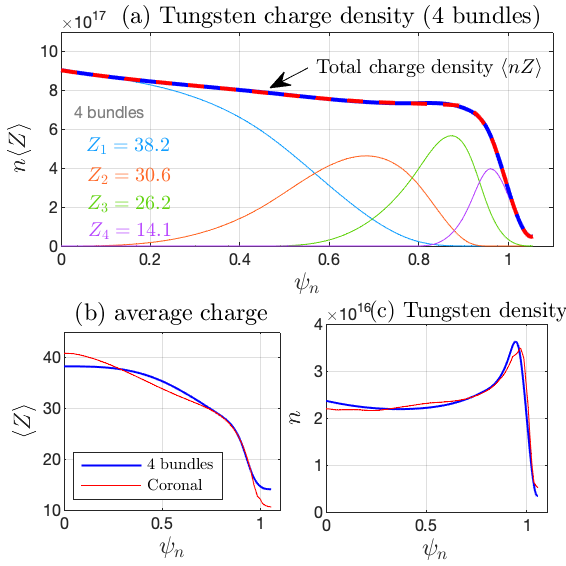}
      \caption{Modeling the multi-ionization states of tungsten with bundles. Subplot (a) shows the modeled charge density in thick blue line. The thin colored curved underneath represent the four bundles we used to model this charge density. Their charge is specified in the figure. Subplot (b) is the average charge $\langle Z\rangle$ modeled accurately with nine finite element (red) and modeled more approximately with four bundles (blue). In subplot (c) the overall tungsten density of the bundles (blue) is compared to the one modeled more accurately with the nine finite-elements. The gathering of finite-elements in bundles allows to decrease the number of gyrokinetic tungsten ions, thus, decreasing the cost of the simulations. }
\label{fig:modeled_4bundles}
\end{figure}

As mentioned in the introduction, our gyrokinetic bundling technique differs from the one employed in fluid codes~\cite{Strachan2011}. {For instance, a fluid entity can have a radial profile of electric charge $\langle Z\rangle$, but a gyrokinetic species has a fixed electric charge Z. The variation of the average electric charge, see Fig.~
\ref{fig:coronal_approximation}(b), is thus modeled with several gyrokinetic bundles of different electric charge $Z_j$ and different radial density profiles $n_j$ that lead to the averaged tungsten  charge $\langle Z\rangle=\langle nZ\rangle/n$ with
\begin{equation}
    \langle nZ\rangle=\sum_jZ_jn_j,
    \label{eq:total_tungsten_charge}
\end{equation}
and
\begin{equation}
    n=\sum_j n_j,
    \label{eq:total_tungsten_density}
\end{equation}
where $j$ labels the different tungsten bundles, see Fig.~\ref{fig:modeled_4bundles}(a).
} 

To compute the initial radial profile of density $n_j$ and the charge $Z_j$ of each bundle $j$, we prescribe an initial radial distribution of tungsten charge density $\langle n Z\rangle$ and we compute the radial profile of average charge $\langle Z \rangle$ from a Coronal approximation and electron temperature, see Fig.~\ref{fig:coronal_approximation}. The profile $\langle Z\rangle (T_e)$ in Fig.~\ref{fig:coronal_approximation} is inspired from Ref.~\cite{Demura15}. 

Once these initial profiles have been defined, we initialize the tungsten bundles from the solution of a linear system where the tungsten particle density is represented with finite elements, such that $n (\psi_n)=\sum_i c_i \Lambda_i(\psi_n)$ and $\langle nZ \rangle(\psi_n)=\sum_i c_i Z_i\Lambda_i(\psi_n)$ with $\Lambda_i$ a 1D finite element (here we use quadratic b-splines see Ref.~\cite{dominski18}). To determine the coefficients $c_i$, we trivially solve the linear system 
\begin{equation}
\begin{cases}
c_i=(M^{-1})_{ij} \int d\psi\,  \Lambda_j(\psi)\,\frac{\langle nZ\rangle(\psi)}{\langle Z\rangle(\psi)}, \\
c_iZ_i=(M^{-1})_{ij} \int d\psi\,  \Lambda_j(\psi)\,\langle nZ\rangle(\psi),
\end{cases}
\label{eq:bundlesystem}
\end{equation}
where $M_{ij}=\int d\psi\, \Lambda_i(\psi) \Lambda_j(\psi)$ is the mass matrix. The radial grid on which the finite elements are defined is irregular such that we have narrower intervals in the region of the pedestal to better model the steep gradient of density. 

To save computational resources, each bundle is either a finite-element or a ``sum'' of finite-elements. The ``summing'' of finite-elements is done in a way that conserves the plasma quasi-neutrality, but allows some inaccuracy in the average density and the charge $\langle Z \rangle$, as shown in Fig.~\ref{fig:modeled_4bundles}(b) and (c).  This choice is made to reduce the cost of a simulation, by reducing the total number of species. The bundles used in Sec.~\ref{sec:multiW_results} of this paper are shown in Fig.~\ref{fig:modeled_4bundles}. Their respective and total charge density are plotted in subplot (a). The total tungsten average charge is plotted in (b) and the total tungsten particle density is plotted in (c). 

\section{An H-mode plasma model in JET Geometry}
\label{sec:JETcase}
\begin{figure}
      \includegraphics[width=8.5cm]{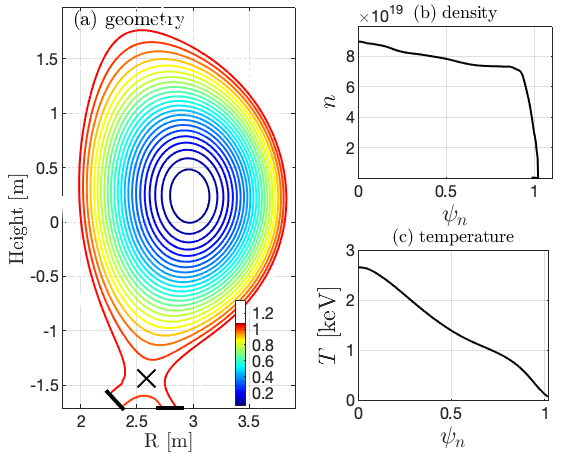}
      \caption{Model plasma in JET geometry - The magnetic equilibrium is illustrated in subplot (a) where the magnetic surfaces $\psi$ are plotted with colors. The density and temperature profiles, including an edge pedestal near $\psi_n>0.8$, are plotted in (b) and (c), respectively. In all our simulations $T_i=T_e=T_{\rm w}$.}
      \label{fig:JET_eD_vs_eDW_equilibirum_poster}
\end{figure}

In this paper, we consider a low-density model H-mode plasma based on a JET carbon-wall discharge designed to optimize the temperature-gradient screening effect, which was already used in Ref.~\citenum{Hager16b}. As an illustration of the magnetic geometry, the magnetic surfaces and divertor of this JET geometry are plotted in Fig.~\ref{fig:JET_eD_vs_eDW_equilibirum_poster} subplot (a), and the modeled profiles of density and temperature are plotted in subplot (b) and (c), respectively.  {Compared to more typical JET-ILW profiles, the temperature pedestal is not so pronounced, which affects the distribution of tungsten ionization states, and the density at the pedestal is low, which affect the collisionality in the pedestal.} More realistic ionization and pedestal conditions are discussed in Ref.~\cite{Angioni14}. The position of the X-point is shown in Fig.~\ref{fig:JET_eD_vs_eDW_equilibirum_poster} (a), together with the position of the divertor surfaces where the open field lines end. Sputtering source of tungsten particles from these plates is not implemented.  Tungsten ions are assumed to exist at the simulation initialization.

Several simulations are presented in the present paper. They are either composed of a single tungsten ion species (single-W) or of multiple tungsten species with different ionization states (multi-W). The multi-W simulations model the tungsten with bundles, as describe in ~\ref{sec:multiple_ionization}. In all these XGCa simulations, multi-species nonlinear Fokker-Planck-Landau collisions are included,  electrons are drift-kinetic and all ions are gyrokinetic. The real physical masses are used. In all figures, the simulation results are shown after $1ms$ when the pure plasma typically reaches a quasi-steady neoclassical state, see Ref.~\cite{Hager2019} where the saturation time of this model plasma is discussed in details. After the quasi-state neoclassical state is reached, plasma profiles relax slowly at the neoclassical transport rate. After $2ms$ there is no noticeable change in the transport fluxes. Longer simulations could bring us to a true neoclassical steady state, but would require switching on the other XGC capabilities: the heat, torque and particle sources with neutral recycling and atomic interactions.

All species are initialized with the same profile of temperature, $T_e(\psi)=T_D(\psi)=T_{\rm w}(\psi)$ at $t=0$. The charge density profiles of deuterium, $n_D$, and tungsten, $\langle Zn \rangle$, are initialized to be constant fractions of the initial electron charge density while ensuring quasi-neutrality. For instance, in all simulations presented in this work, the initial deuterium density is chosen to represent $99\%$ of the electron density at all positions, $n_D=0.99\times n_e$ at $t=0$. The total tungsten charge represents the remaining $1\%$ of the electron charge density. The density is adapted to the tungsten species charge $Z$. For $Z=40$, the tungsten fraction relative to the electron density is $n_{\rm w}/n_e=2.5\times10^{-4}$, a number which may be ~10$\times$ higher than ITER but relevant to physics studies.

\begin{figure}
      \includegraphics[width=8.5cm]{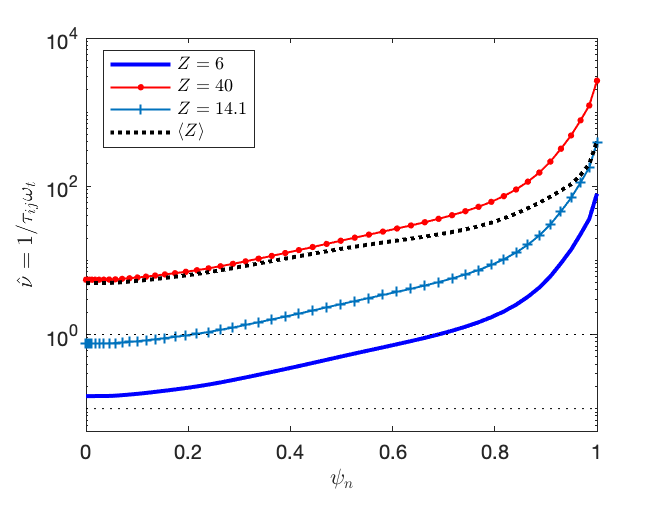}
      \caption{Tungsten collisionality $\hat\nu$ in the model JET plasma introduced in Fig.~\ref{fig:JET_eD_vs_eDW_equilibirum_poster}, for three different charge states $Z=6, 14.1, 40$ and for the profile of average charge $\langle Z\rangle$. Tungsten-deuterium collisions are more important than tungsten-tungsten or tungsten-electrons collisions. See Ref.~\cite{Dominski19a} for details on the collisionality calculation.}
      \label{fig:collisionality_profile}
\end{figure}

In the present simulations, deuterium particles are in the banana-plateau (BP) regime inside the separatrix and in the collisional Pfirsch-Schl\"uter (PS) regime in the SOL. Tungsten is so heavy and so electrically charged that it is in general in the PS regime at all radial positions. Only very low-Z ($Z\sim 6$) tungsten particles are near the transition between BP and PS inside the pedestal top. Profiles of representative collisionality are shown in Fig.~\ref{fig:collisionality_profile} for three charge states of tungsten ions and for the averaged charge $\langle Z\rangle$ modeled with bundles.

No source of torque is used, but a parallel flow naturally arises in our total-f gyrokinetic model and satisfies the radial force balance. The parallel flow speed, $U_\parallel$, of tungsten impurities is measured to be higher near the edge. Its Mach number relative to the background sound speed, $M=U_\parallel/\sqrt{T_e/m_D}$, is up to $\simeq10\%$ in the pedestal and up to $40\%$ in the SOL. As we do not use any source of external torque in the plasma, the observed parallel flow in the pedestal inside the separatrix is mostly a consequence of the ion orbit losses near the boundary and of our initial condition. Indeed, the particle distribution function $f$ is initialized as a Maxwellian without mean flow and its relaxation leads to the appearance of parallel flow with a polarization of the plasma by conservation of the toroidal canonical angular momentum, see Ref.~\cite{Ku04,Chang08}.

Let us end this section by discussing the length scales and ordering of interest. The Larmor radius 
\[
\rho_i=\frac{\sqrt{Tm_i}}{Z_ieB}
\]
is proportional to $\sqrt{m_i}/Z_i$ such that the tungsten Larmor radius increases as $Z$ decreases: $\rho_{\rm w}(Z=40)\simeq0.25\rho_D$, $\rho_{\rm w}(Z=14)\simeq0.7\rho_D$, and $\rho_{\rm w}(Z=6)\simeq1.6\rho_D$. It can be immediately noticed that the usual small banana width orderings used in the analytic theories for low Z tungsten particles could be problematic in the narrow pedestal region when their banana orbit width can be as wide as the pedestal gradient scale length. A non-local total-f gyrokinetic simulation can provide more accurate physics results.

\section{Simulations with a single tungsten impurity species}
\label{sec:singleW_results}

In this section, we study plasmas with the tungsten impurity in a single ionization state. We will compare two simulations: one using a tungsten species with $Z=6$ and a second one using a tungsten species with $Z=40$.

\subsection{Inward and outward fluxes of tungsten and associated asymmetries}
\begin{figure}
      \includegraphics[width=8.5cm]{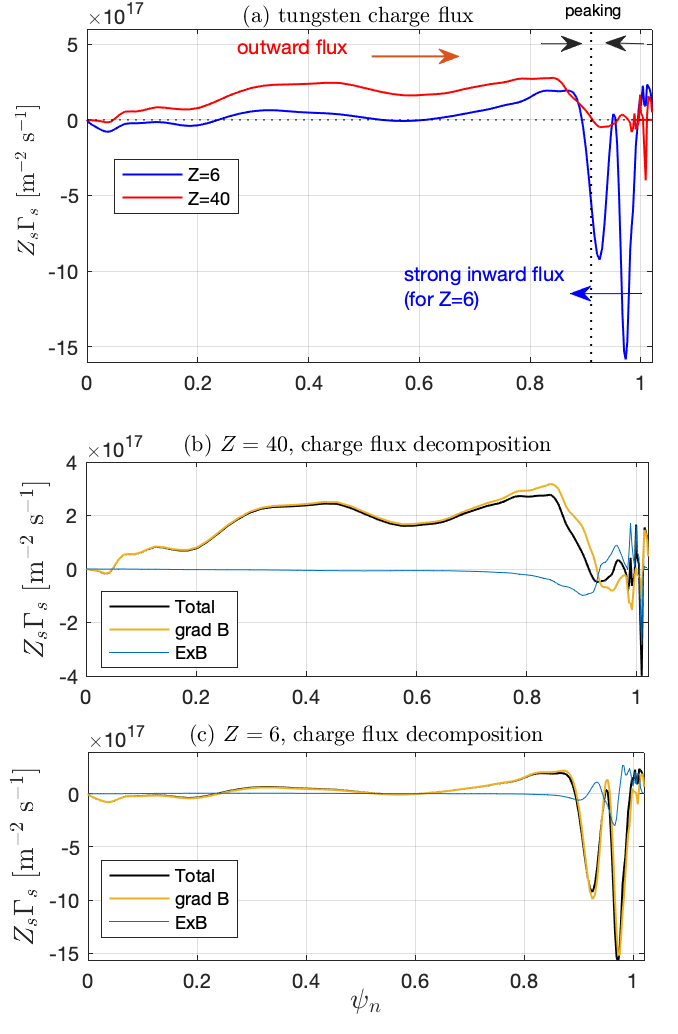}
      \caption{Comparing the radial flux of tungsten charge in simulations carried out with different tungsten ionization state $Z$. (a) Radial profile of the tungsten charge flux for single-W simulation with $Z=40$ in red and $Z=6$ in blue. These radial profiles represent the charge flux split into its $E\times B$ and grad B contributions for both (b) $Z=40$ and (c) $Z=6$ simulations. The grad B contributions dominate the $E\times B$ ones in both cases. }
      \label{fig:fluxes_tungsten_decomposition_ExB_gradB}
\end{figure}

The radial flux of tungsten charge, $Z_{\rm w}\Gamma_{\rm w}$, is illustrated in Fig.~\ref{fig:fluxes_tungsten_decomposition_ExB_gradB}(a) for both simulations. In the simulation with $Z=40$ (red), the charge flux is directed outward in the core and weakly inward in the edge. This naturally results in an impurity peaking near the pedestal top, where the flux is zero. Conversely, in the simulation with $Z=6$ (blue), the charge flux is notably inward in the pedestal, shifts outward after the pedestal top (around $\psi_n \simeq 0.7 - 0.85$), and approaches zero in the core. In both these low-Z and high-Z cases, the magnitude of the tungsten charge flux is sufficiently large at certain locations to significantly impact the main-ion charge flux, particularly in the core for the simulation with $Z = 40$. The influence of tungsten on main-ion fluxes will be explored in Sec.~\ref{sec:ionfluxes}. A robust correlation between deuterium and tungsten fluxes is anticipated from neoclassical theory ($\Gamma_D\simeq-Z_{\rm w}\Gamma_{\rm w}$), as electrons are too light to contribute significantly to momentum conservation in inter-species Coulomb collisions

In these simulations, the dominant contribution to the particle flux is from the grad-B drift which consequently determines the overall trend of the total flux, see yellow and black cures in Fig.~\ref{fig:fluxes_tungsten_decomposition_ExB_gradB}. Conversely, the $E\times B$ contribution (blue) becomes more pronounced in the pedestal and SOL region. The grad-B and $E\times B$ contributions to the charge flux are defined by
\begin{equation}
    \begin{cases}
    Z\Gamma_{\nabla B}=Z\left\langle\nabla\psi\cdot\int d\mathbf{v}\, \mathbf{v}_{\nabla B} f\right\rangle \\ 
    Z\Gamma_{E\times B}=Z\left\langle\nabla\psi\cdot\int d\mathbf{v}\, \mathbf{v}_{E\times B} f\right\rangle
    \end{cases}.
\end{equation}
The total charge flux, representing the particle flux $\Gamma$ weighted by the charge number $Z$, is given by the sum of the grad-B and $E\times B$ contributions: $Z\Gamma=Z(\Gamma_{\nabla B}+\Gamma_{E\times B})$.

\begin{figure}
      \includegraphics[width=8cm]{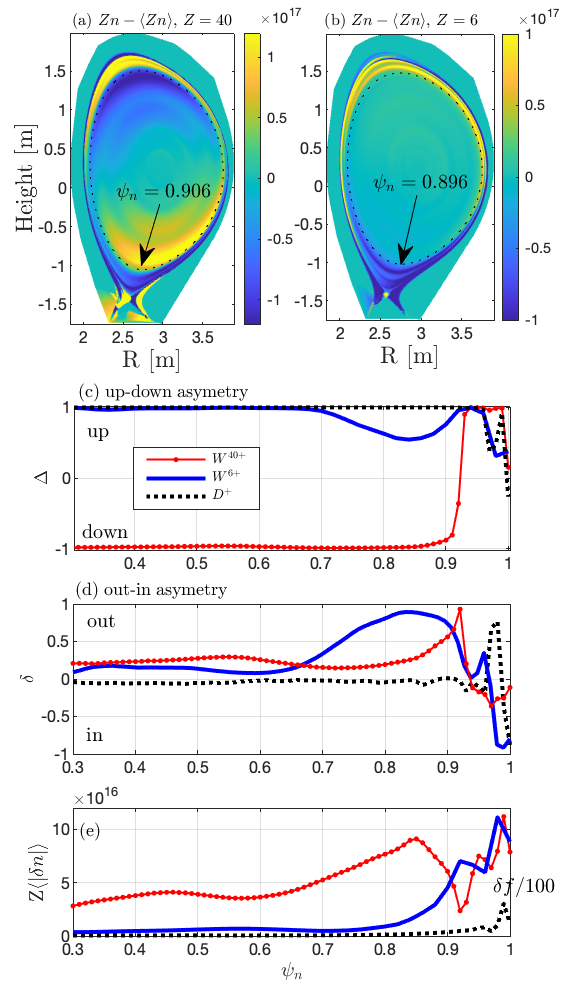}
      \caption{Asymmetries of the tungsten density in simulations carried out with $Z=6$ (blue) and $Z=40$ (red with dot markers). The 2D maps of poloidal asymmetry is computed with $\delta n=n-\langle n \rangle_{FS}$ for both the (a) $Z=6$ and (b) $Z=40$ simulations. From these 2D maps, we compute: (c) the up-down asymmetry, (d) the in-out asymmetry, and (e) the amplitude of the poloidal density perturbation $\langle|\delta n|\rangle$. Note that $n\simeq n_0+\langle|\delta n|\rangle (\delta\cos\theta+\Delta\sin\theta )$.
      }
      \label{fig:tungsten_2Dmap_and_asymmetries}
\end{figure}

The direction of the particle flux correlates significantly with the up-down poloidal asymmetries~\cite{Fulop99,Angioni14b} of the tungsten density, shown in Fig.~\ref{fig:tungsten_2Dmap_and_asymmetries} (a,b,c). In a regime of strong collisionality where the parallel friction force dominates the dynamics and the grad-B drift points downward, an accumulation at the top will be pushed downward from the top to the core of the plasma and translates to an inward flux of particles~\cite{Fulop99,Angioni14b}. Respectively, an accumulation at the bottom leads to a downward flux towards the edge region and translates into an outward flux. This trend is observed for $Z=40$ at all radial positions and for $Z=6$ near the edge where those impurities are in the strong collisional regime. It is not the case in the core of the $Z=6$ simulation where it is in a less collisional regime, see collisionality in Fig.~\ref{fig:collisionality_profile}. Note that 2D asymmetries could also affect the contribution of the $E\times B$ drift to the radial flux, but this $E\times B$ drift flux appears to be secondary in our two simulations. Let us recall that we do not include turbulent $E\times B$ transport.

In these simulations, we do not have a source of torque but we observe in-out asymmetries, see Fig.~\ref{fig:tungsten_2Dmap_and_asymmetries}(d), that result from the moderate rotation caused by our initial conditions. As mentioned in Sec.~\ref{sec:JETcase}, the rotation in the core is related to the relaxation of $f_0$ initiated as a local Maxwellian and the rotation in the edge is related to ion-orbit loss effects near the separatrix. Studying the role of in-out asymmetries in realistic JET plasma would require the inclusion of toroidal torque and larger rotation, and also the effects of ICRH {which can dominate the drive for in-out asymmetries~\cite{Angioni14}. Nonetheless, given that ITER will operate with mild rotation and electron heating (gyrotrons), our model plasma configuration is of interest.

\subsection{Temperature screening effect and asymmetries}
In this section, we discuss the temperature screening effect and its correlation to the poloidal asymmetries. Our model plasma is well suited for this effect, as the temperature gradient is substantially larger than the density gradient, as illustrated in Fig.~\ref{fig:temperature_screening}(b). 

From neoclassical theory~\cite{Fajardo22,Fajardo23}, considering tungsten as a single species (no bundles), the particle flux can be estimated with 
\begin{equation}
    \Gamma/nZ = -D(\nabla\ln{n})^*
\label{eq:gamma_neo}
\end{equation}
where
\begin{equation}
(\nabla\ln n)^*=\left(1-\frac{1}{Z}\right)\nabla\ln{n}+\frac{H}{K}\nabla\ln{T},
\label{eq:gradnstar}
\end{equation}
and $K=ZD$. For the sake of simplicity, this expression is simplified as all ions are initialized with the same radial profiles of logarithmic gradients $\nabla\ln{n}$ and $\nabla\ln{T}$ and we neglect the small relaxation occurring during the simulations. 

The radial profile of $(\nabla\ln n)^*$ is estimated in Fig.~\ref{fig:temperature_screening}(a) for three values of the temperature screening (TSC) coefficient ${H}/{K}$. Results show that the profile of flux observed in our gyrokinetic simulations with $Z=40$ in Fig.~\ref{fig:fluxes_tungsten_decomposition_ExB_gradB}(a) match the profile observed in the case of a favorable screening effect with ${H}/{K}\simeq-0.5$ (blue). The particle flux is outward in the core, even more outward near the pedestal top, and somewhat inward near the separatrix. In comparison when neglecting the screening effect (red) the particle flux is inward in the whole plasma volume, from the separatrix down to the magnetic axis. The flux measured in the simulation with $Z=6$ seems to be in between these two cases: a weak outgoing flux near pedestal top and a strong inward flux near the separatrix.
 \begin{figure}
      \includegraphics[width=8.5cm]{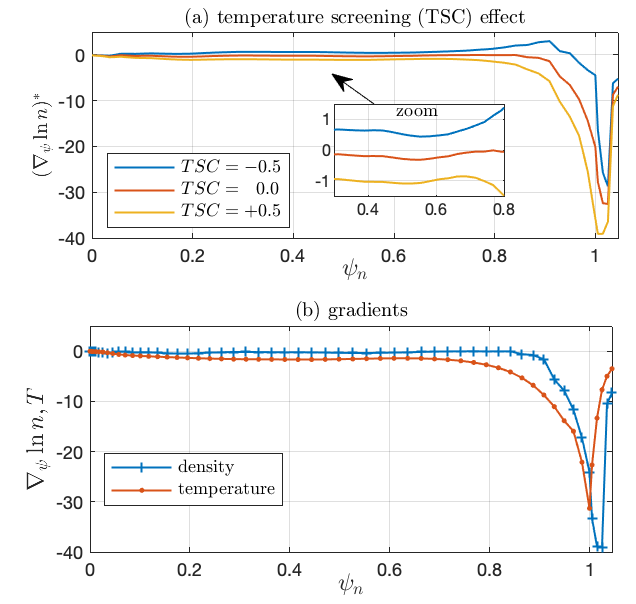}
      \caption{Temperature screening (TSC) effects highlighted by evaluating (a) the effective gradient with three different values of $H/K=-0.5,\ 0,\ +0.5$. This effective gradient is computed from the density and temperature gradients shown in (b). It is remarkable that the temperature gradient is larger than the density gradient at all positions before the separatrix, suggesting a significant role of temperature screening effects.
      }
      \label{fig:temperature_screening}
\end{figure}

Asymmetries can influence the impurity flux as highlighted by coefficients $\mathcal{C}_G$ and $\mathcal{C}_U$ used to estimate\cite{Fajardo22,Fajardo23} the Pfirsch-Schl\"{u}ter impurity diffusion 
\begin{equation}
    D=q^2\rho^2\nu\frac{\mathcal{C}_G}{2\epsilon^{2}}
\end{equation}
and screening factor
\begin{equation}
    \frac{H}{K}
    =-\left[  \frac{1}{Z}+(C_0^z-1)\right] +\frac{\mathcal{C}_U}{\mathcal{C}_G}\left(C_0^z+k_i\right),
\end{equation}
where $Z_i=1$ has been used for the main ion~\cite{Fajardo22},
\begin{equation}
    \mathcal{C}_G=\langle n/b^2 \rangle - 1/\langle b^2/n\rangle,
\end{equation}
\begin{equation}
    \mathcal{C}_U=\langle n/N \rangle - \langle b^2/N\rangle/\langle b^2/n\rangle,
\end{equation}
$n=n_{\rm w}/\langle n_{\rm w}\rangle$, $b^2=B^2/\langle B^2\rangle$, $N=n_D/\langle n_D\rangle$, $C_0^z\simeq1.5$ is a friction coefficient, and $k_i$ is the ion flow coefficient. 

The asymmetry coefficients $\mathcal{C}_G$ and $\mathcal{C}_U/\mathcal{C}_G$ are shown in Fig.~\ref{fig:asymetrie_infleunce_on_flux} (a) and (b), respectively. 
The asymmetry factor $\mathcal{C}_U/\mathcal{C}_G$, which influences the screening factor, varies significantly between these two simulations. For instance, it increases by up to a factor 20 in the core of the simulation using $Z=40$ instead of $6$. Moreover, it is remarkable that the profile of the asymmetry factor $\mathcal{C}_U/\mathcal{C}_G$ qualitatively matches the profiles of particle flux from Fig.~\ref{fig:fluxes_tungsten_decomposition_ExB_gradB} (a). In particular it captures the difference between the simulations carried out with $Z=6$ and $40$, even the almost zero particle flux near the separatrix with $Z=40$. In comparison, the asymmetries only mildly influence the diffusion coefficient $D$, see Fig.~\ref{fig:asymetrie_infleunce_on_flux} (a), and do not explain much of the difference observed between the particle flux measured in $Z=6$ and $40$ simulations. These observations suggest that asymmetries affect the temperature screening effect thus contributing to the difference between our two simulations. Quantitatively $C_0^z+k_i\simeq\pm1$ does not entirely explains the large flux modulation we observe in our simulation, but we made various approximations when estimating these quantities. A theoretical analysis of the role of asymmetries on the screening factor and on other neglected components will be studied in a future work.

 \begin{figure}
      \includegraphics[width=8.5cm]{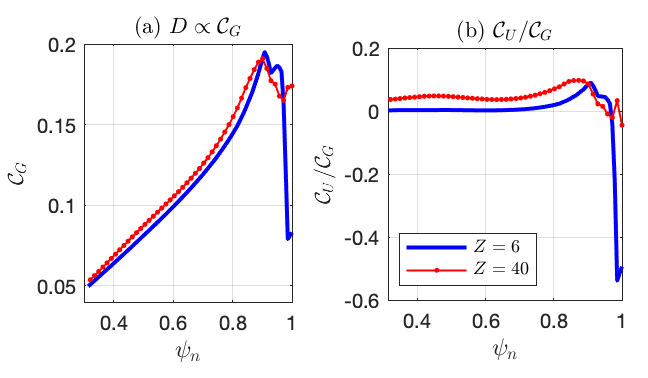}
      \caption{Asymmetries coefficient influencing the (a) diffusion and (b) screening factor. Simulation with $Z=6$ in blue and $Z=40$ in red with dot markers. Data is computed from asymmetries shown in  Fig.~\ref{fig:tungsten_2Dmap_and_asymmetries}.
      }
      \label{fig:asymetrie_infleunce_on_flux}
\end{figure}
}

\subsection{Drive for up-down asymmetries of tungsten density}
\label{sec:asym_singleW}
Tungsten $Z=40$ is in the same high collisional regime at all radial positions, but its up-down asymmetry changes direction near the separatrix. Indeed, tungsten $Z=40$ accumulates at the bottom before $\psi_n\simeq0.92$ and at the top after, see Fig.~\ref{fig:tungsten_2Dmap_and_asymmetries} (c). 

This reversal of the up-down density asymmetry near the edge is understood to be due to a reversal of the direction of the friction force, $R_\|$, and to an increase of the parallel electric field , $E_\|$. In the core $R_\|$ pushes tungsten at the bottom and in the edge both $R_\|$ and $E_\|$ push tungsten at the top, as we will discuss in more details in the simulation with bundles of tungsten in Sec.~\ref{sec:bundle_flux_reversal}. Let us recall, that in the high collisional regime, the up-down asymmetries are driven by the parallel electric field and parallel friction force\cite{Helander98,Dominski19a}
\begin{equation}
\nabla_\parallel \ln{n_w}\simeq\frac{eZ}{T_{\rm w}}E_\parallel+\frac{1}{n_{\rm w}T_{\rm w}}R_{\parallel,w},
\label{eq:momentum}
\end{equation}
and at highest collisionality~\cite{Dominski19a} ($Z=40$) the parallel electric field becomes negligible, leading to
\[
    \nabla_\|\ln{n_{\rm w}} \simeq  \frac{1}{n_{\rm w}T}\,R_\|.
\]

Tungsten $Z=6$ is also highly collisional in the edge where its asymmetry accumulates at the top like for $Z=40$. Nonetheless in the core, tungsten $Z=6$ is less collisional ($\sim$plateau) than tungsten $Z=40$ and it has an up-down asymmetry of opposite direction.  Indeed, tungsten $Z=6$ is less sensitive to the friction force in the core because of its lower collisionality there, which explains that the parallel electric field (and viscous force) can become the dominant driver of up-down asymmetry, see Ref.~\cite{Dominski19a}. 

\section{JET-model plasma simulations with multiple ionization states of tungsten}
\label{sec:multiW_results}
In this section, we study a JET-model plasma including multiple ionization states of tungsten bundled into four charge states, see bundle description in Sec.~\ref{sec:multiple_ionization}. Our simulations capture the quasi-steady neoclassical transport fluxes, but they do not capture the relaxation of profiles consistently with source terms over the transport time scale. As a result, the final radial profiles depicted in Fig. 9(a) exhibit a small deviation from the initial profiles shown in Fig. 2(a).

The discussion will focus on the radial transport carried out by the bundles with $Z=26.2$ and $Z=14.1$ because they are the dominant bundles in the edge. In the core, the transport of bundles $Z=38.1$ and $30.6$ are found to be similar to the one of the simulation carried out with the single tungsten ion $Z=40$.

\subsection{Radial profile of particle fluxes modeled with bundles}
 \begin{figure}
     \includegraphics[width=7.5cm]{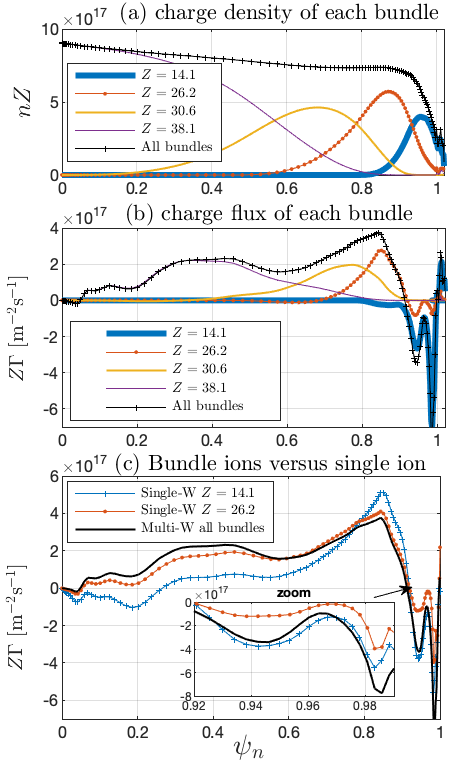}
     \caption{For each tungsten bundle, we show the radial profile of (a)  charge density and of (b) radial flux of charge measured at the end of the simulation. (c) Comparison of the multi-W simulation with two single tungsten ion simulation (blue and red lines with dot markers).
     }
     \label{fig:fluxes_impurities_multiW}
\end{figure}

 \begin{figure}
     \includegraphics[width=7.5cm]{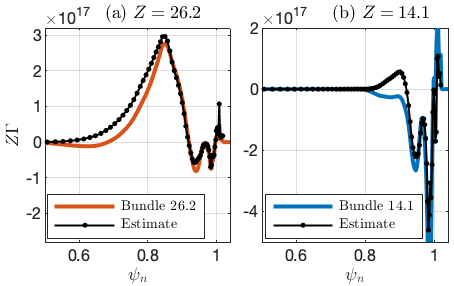}
     \caption{Estimate of the bundle flux from a simulation carried out with a single ion simulation of same charge. The estimate is computed using the relation $\Gamma_{\rm B} \simeq \frac{n_{\rm B}}{n_{\rm w}} \Gamma_{\rm w}$, see Eq.~\eqref{eq:renormalizedflux}, in order to compensate for the different profiles density.
     }
     \label{fig:single_vs_bundle_estimate}
\end{figure}

The total radial flux of tungsten ions (black) is shown in Fig.~\ref{fig:fluxes_impurities_multiW} (b). The contribution of each bundle is shown in color. The bundle $Z=14.1$ with the lowest charge-state moves inward across the separatrix surface into the pedestal top, while the bundle $Z=26.2$ with higher charge state moves out toward the pedestal top. In the core, the bundles $Z=38.1$ and $Z=30.6$ sustain an outgoing flux of particles consistent with our previous single ion simulation $Z=40$.  

In Fig.~\ref{fig:fluxes_impurities_multiW} subplot (c) we present a comparison of this multi-W simulation with the single-W simulations by carrying out two additional simulations. One is using the charge state $Z=26.2$ and the second one is using the charge state $14.1$. The single-W simulation using $Z=14.1$ underestimates the outgoing flux in the core and overestimates the outgoing flux in the pedestal top. The single-W simulation using $Z=26.2$ is close to the multi-ions simulation in the core but strongly underestimates (by $\lesssim50\%$)  the inward flux sustaining the penetration of tungsten in the pedestal ($\psi_n\gtrsim0.9$), see the zoom in (c).
      
The \textit{local} transport trend can still be deduced from a single tungsten ion simulation if the charge is relevant to the region of interest with $Z\simeq\langle Z\rangle$, as the local dynamics of each bundles appears to be dominated by its interaction with the main plasma and with itself. Indeed, the transport of each bundle is consistent with the single tungsten simulation using the same charge number, see Fig.~\ref{fig:single_vs_bundle_estimate} (a) and (b) where the tungsten flux estimates (black) are obtained by re-normalization of the flux computed in single-ion simulations. This re-normalization consists in multiplying the flux of a single tungsten species, $\Gamma_{\rm w}$, by the ratio of the density profiles to obtain an estimate of the tungsten bundle flux, $\Gamma_{\rm B}$, expressed as follows
\begin{equation}
\Gamma_{\rm B} \simeq \frac{n_{\rm B}}{n_{\rm w}} \Gamma_{\rm w},
\label{eq:renormalizedflux}
\end{equation} 
where $n_{\rm B}$ is the bundle density profile and $n_{\rm w}$ is the single tungsten density profile.

A systematic difference between estimated and measured bundle fluxes is visible on the left hand side of the flux profile in Fig.~\ref{fig:fluxes_impurities_multiW} subplots (a) and (b). At these locations, the bundle flux is lower and sometimes even of opposite direction (inward) than in a single ion simulation. This can be explained by the the particular profile of density of each bundle. Indeed, at these locations the bundle density gradient is of opposite sign compared to the total tungsten density gradient thus contributing to a diffusion in the opposite direction. Second, the selected bundle density is much smaller than the one of the contiguous bundle and potentially reacts to collision with these other bundles.

 \begin{figure}
      \includegraphics[width=8cm]{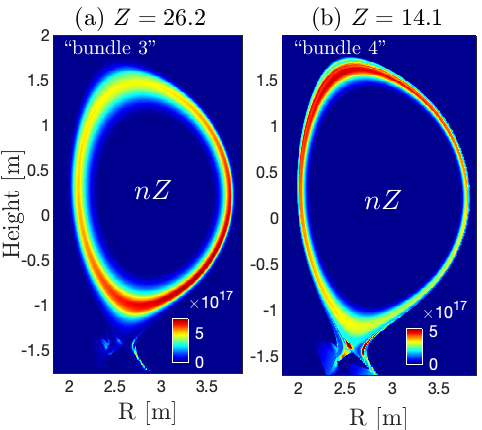}
      \includegraphics[width=8cm]{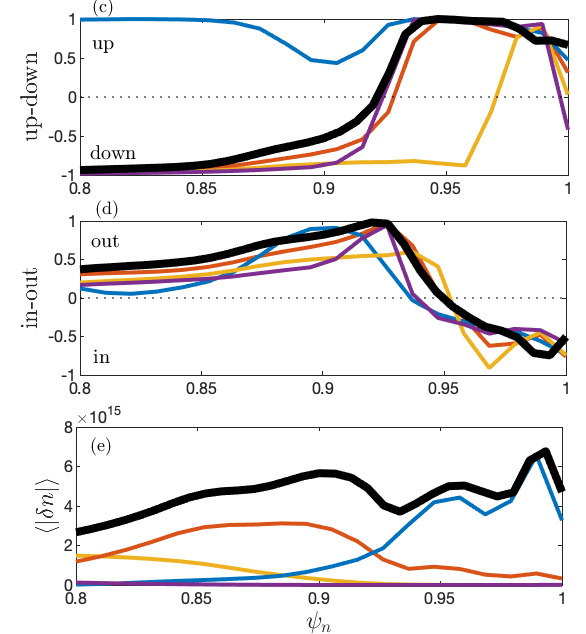}
      \caption{Density organization and radial fluxes of the bundle $Z = 26.2$ and bundle $Z = 14.1$ in and around the pedestal. Subplots (a) and (b) show the 2D density of tungsten these bundles. Subplot (c) measures the up-down asymmetry, (d) the in-out asymmetry, and (e) the amplitude of the poloidal density perturbation. Same definitions than for Fig.6. Color lines represent the bundles and the thick black line is the total tungsten, same color definition than in Fig. 9 (a).
      }
      \label{fig:Parallel_friction_force_influence}
\end{figure}
  
\begin{figure}
      \includegraphics[width=8cm]{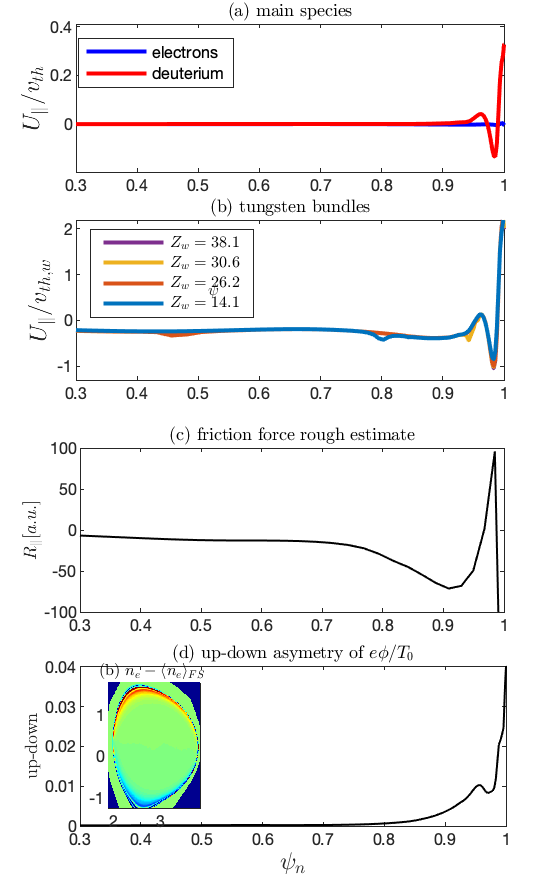}
      \caption{Parallel flow of (a) deuterium and electrons and of (b) tungsten bundles. (c) Friction force estimate based on Eq.~\eqref{eq:frictionforce} with the profile of charge  $\langle Z\rangle$. (d) Up-down asymmetry of the electrostatic field with a zoom on the related electron 2D map of density. $U_\|$ is normalized to the respective species thermal velocity.}
      \label{fig:new_parallel_flow}
\end{figure}

\subsection{Asymmetries of the tungsten modeled with bundles}
      \label{sec:bundle_asymetries}
The poloidal asymmetries of charge density, which are source of neoclassical radial transport~\cite{Fulop99,Angioni14b} in tokamaks, are computed for bundles $Z=26.2$ and $14.1$ and shown in Fig.~\ref{fig:Parallel_friction_force_influence}. The up-down, $\Delta$, and in-out, $\delta$, asymmetries of tungsten density are shown in subplots (c) and (d), respectively. These asymmetries are shown for each bundle density in color and for the total tungsten density in black. The total density map is reconstructed using the following equation
\begin{equation}
n\simeq n_0+\langle|\delta n|\rangle (\delta\cos\theta+\Delta\sin\theta ),
\end{equation}
where $\langle|\delta n|\rangle$ is the amplitude of the poloidal perturbation shown in (e) and $n_0$ is the background density.

The total tungsten density (black curves) exhibits a downward accumulation in the core and an upward accumulation in the edge, see subplot (c). The transition from downward to upward asymmetry is sharp and occurs near $\psi_n=0.92$. This reversal of the up-down asymmetries coincides with a reversal of the direction of the particle flux. With the grad-B drift pointing downward, an accumulation of tungsten at the top sustains an inward flux of particle, whereas an accumulation at the bottom sustains an outward flux of particles.

In comparison, in-out asymmetries are weaker than up-down asymmetries but they are still present, see (d). In particular, there is a peak of outward accumulation where the up-down asymmetry is zero (near $\psi_n=0.92$). This can be understood by the fact that at this location, the drive for up-down asymmetries cancels and, therefore, the weaker drive for an outward accumulation becomes locally dominant. It is also remarkable that this in-out asymmetry is outward up to $\psi_n=0.95$ and then inward after.

The internal organization of the tungsten asymmetries shows some differences between the different bundles. In particular, the low-Z bundle ($Z=14.1$) has an up-down asymmetry in the core in the opposite direction from other bundles. The up-down asymmetry difference can be explained by its low collisionality that weakens the importance of the parallel friction force and strengthen the importance of other drives including the parallel electric field and viscous force~\cite{Helander98,Dominski19a}. On the other hand, in-out asymmetries are similar for all tungsten bundles, see Fig.~\ref{fig:Parallel_friction_force_influence}(d). This is understood by the fact that they have a similar parallel flow, see Fig.~\ref{fig:new_parallel_flow} (b). 

\subsection{Effect of the X-point ion orbit loss physics}

\label{sec:bundle_flux_reversal}
In this subsection, we analyze how the X-point~\cite{Chang02} ion-orbit losses modulate the ions parallel flows $U_\|$, see Fig.~\ref{fig:new_parallel_flow}, near the separatrix thus contributing to the reversal of the up-down asymmetries and particle flux of tungsten. The connection between the two is the friction force which depends on the direction of the parallel flows and can approximately estimated with
\begin{equation}
R_\|\simeq mn\nu(U_{\|{\rm w}}-U_{\|i}),
\label{eq:frictionforce}
\end{equation}
see subplot (c).

The transition of the up-down asymmetry direction near the edge is related to a change in the ions' parallel flows. Up to $\psi_n\simeq0.9$ the deuterium flow is nearly zero and the tungsten flow is negative. Their difference makes the parallel friction force to be negative, $\Delta U_\|= U_{\|{\rm w}}-U_{\|i}<0$, and pushes ions downward. Then, radially outside of this location the deuterium parallel flow grows positive and the tungsten flow increases until becoming positive near $\psi_n=0.95$. Their difference thus becomes positive, $\Delta U_\|>0$, and drives a positive parallel friction force that pushes tungsten ions upward and sustains an inward radial flux of tungsten particles. We recall that up-down asymmetries are shown in Fig.~\ref{fig:Parallel_friction_force_influence} (c) and the particle flux is shown in Fig.~\ref{fig:fluxes_impurities_multiW} (b). 

As discussed in Sec.~\ref{sec:asym_singleW}, up-down asymmetries are also driven upward by the parallel electric field which increases in amplitude near the edge, thus contributing to the up-down asymmetry reversal, see Fig.~\ref{fig:new_parallel_flow}. Also, tungsten is less charged near the edge and becomes more sensitive to this parallel electric field contribution~\cite{Dominski19a} in this plasma configuration ($ZE_\|/R_\|\propto 1/Z$). 
      
\begin{figure}
      \includegraphics[width=7cm]{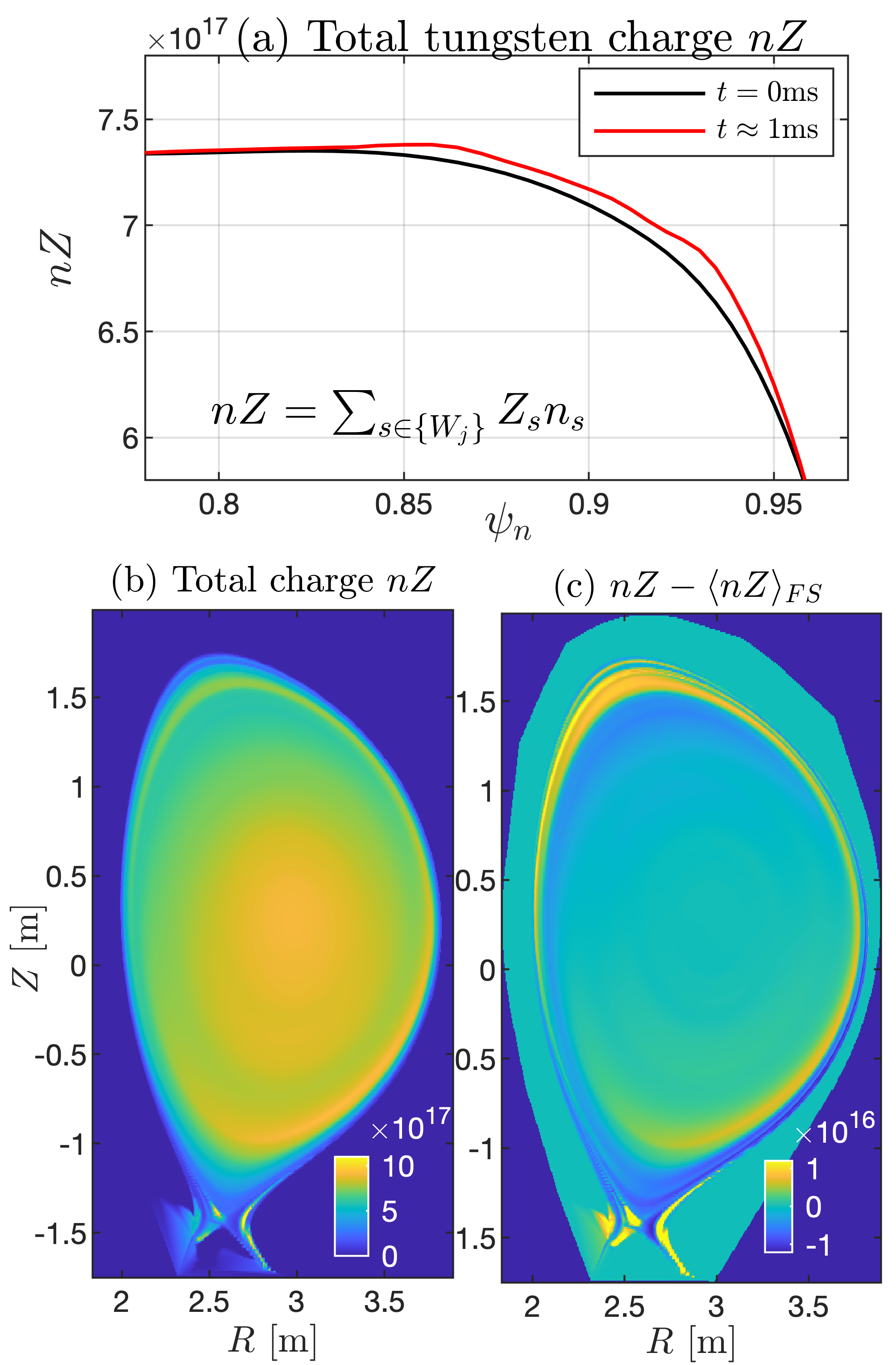}
      \caption{Total tungsten charge accumulation in the pedestal of the multi-tungsten simulation with bundles. The total charge is noted $(Zn)_{\rm w}$ and is equal to $(Zn)_{\rm w}=\sum_{s\in\{W_j\}}Z_sn_s$, where $\{W_j\}$ represents the ensemble of tungsten bundles. The radial profile is shown in (a), the 2D map is shown in (b). In (c), we subtract the zonal flow average to highlight the large asymmetric structure near the pedestal.}
      \label{fig:density_accumulation_multi_ionizations}
\end{figure}
\subsection{Tungsten screening and accumulation in the pedestal top}
One general observation is that the inward motion of the low-Z tungsten from SOL into the pedestal and the outward motion of the higher charge tungsten bundles (screening) toward the pedestal top make the net tungsten charge density to accumulate in the pedestal top area: see Fig.~\ref{fig:density_accumulation_multi_ionizations}(a). This accumulation at the pedestal top occurs in the low field side region, see subplots (b) and (c) of Fig.~\ref{fig:density_accumulation_multi_ionizations} and is consistent with the centrifugal force effect that form spontaneously in our gyrokinetic simulation, see Fig.~\ref{fig:new_parallel_flow}. The pedestal is where the flow is the fastest in our model plasma, by at least an order of magnitude compared to the core region. Finally, even if our modeling differs from rotating JET plasma reported in Ref.~\cite{Kochl18}, it is interesting to point out that in this Ref.~\cite{Kochl18} the tungsten accumulation is similarly observed in the pedestal top region. Note that the contribution of turbulence to rotation effects reported in Ref.~\citenum{Casson15} is not included in our simulations. 

As mentioned in Sec.~\ref{sec:bundle_asymetries}, the tungsten accumulation tends to be outward in the narrow radial layer where the up-down asymmetry reverses its direction and goes from a bottom accumulation to a top accumulation, because within this transition the up-down asymmetries has a zero point where the in-out asymmetry becomes locally dominant.
 
For the purpose of our temperature-gradient screening study, the electron density gradient is modeled to be mild while the ion temperature gradient is steeper.  It will be of interest to vary the ion temperature and electron density profiles in the way relevant to ITER, and to find out how it affects the temperature screening and transport of high-Z tungsten ions. 

\subsection{Influence of tungsten impurities on the radial electric field}
\begin{figure}
      \includegraphics[width=8.5cm]{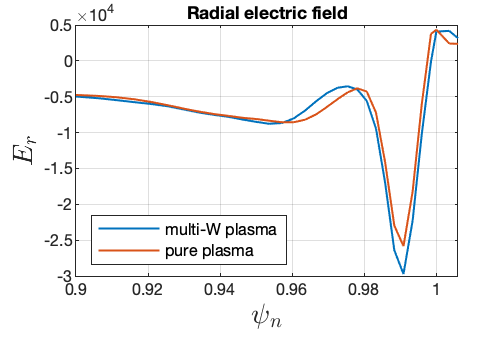}
      \caption{Radial electric field in the pedestal, computed with $E_r=\bf{E}.\nabla \psi/|\nabla \psi|$.}
      \label{fig:electric_field}
      \end{figure}
Given the observed influence of ion-orbit loss on the transport of tungsten in the pedestal, we looked at the influence of the tungsten ions on the radial electric field $E_r$ in the edge region. As can be seen in Fig.~\ref{fig:electric_field}, the modification of $E_r$ from the pure deuterium-electron plasma to the tungsten contaminated plasma at experimentally relevant level is only moderate. Its influence on the plasma turbulence will be studied in the near future.

\section{Influence of tungsten impurities on Deuterium neoclassical fluxes}
\label{sec:ionfluxes}
\begin{figure}
      \includegraphics[width=8.5cm]{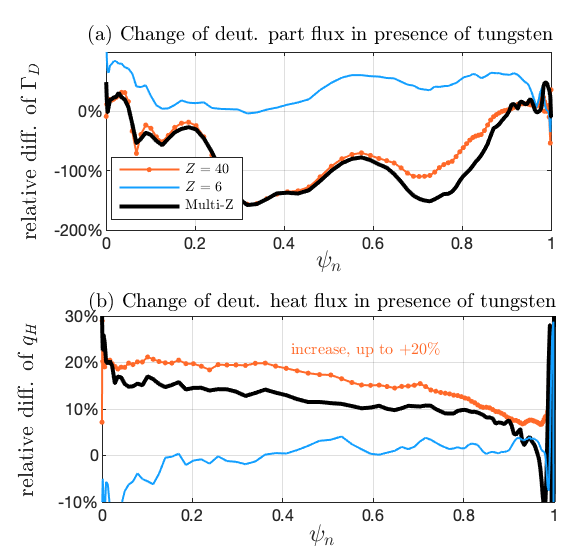}
      \caption{Influence of tungsten impurities on ion neoclassical fluxes. (a) ion particle flux and (b) ion heat flux in three different simulations: single tungsten simulation with $Z=6$ (blue) and $Z=40$ (red), as well as bundled tungsten simulations with $z=\langle Z\rangle^{Coronal}$ (black).}
      \label{fig:JET_eD_vs_eDW_Deut_fluxes}
\end{figure}
In our model plasma with a significant concentration of tungsten, the deuterium fluxes of particles and heat are affected by the presence of tungsten impurities. Fig.~\ref{fig:JET_eD_vs_eDW_Deut_fluxes} shows the tungsten effect on the deuterium {neoclassical} fluxes of particle and heat. In presence of high-Z impurities (red or black in the core) is causing a considerable reduction of the radially outward deuterium particle flux and even a reversal of its direction at some positions. It can also be seen that the influence on the deuterium transport by tungsten ions modeled with the multi-W bundles is in good agreement with the one modeled by a single high-Z tungsten in the core and with the one modeled with a low-Z tungsten in the edge. The largest error made by the simulation using a single high-Z tungsten ion species (red) is on the estimate of the heat flux in the pedestal, after $\psi_n\gtrsim0.9$ in (b). On the other hand, the simulation using a single low-Z tungsten ion species (blue) does not capture the modification of the deuterium flux in the core, but it agrees well with the multi-W simulation in the pedestal. Let us point out that the convective contribution of energy transport is smaller than the diffusive one. We note here that the neoclassical tungsten collisionality, thus transport, increases linearly with Z even for the same tungsten charge density $n_Z Z$ since $\nu\propto n_Z Z^2=(n_Z Z) Z$. 

\begin{figure}
  \includegraphics[width=8cm]{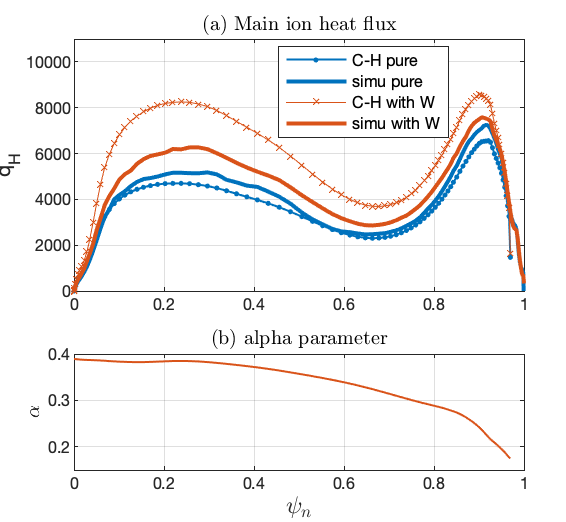}
  \caption{Ion heat flux against Chang-Hinton formula w/o and with the influence of impurities through the $\alpha$ parameter.}
  \label{fig:JET_qH_fluxes_against_CH}
\end{figure}

A comparison of the heat flux of deuterium measured in our simulation and the one predicted by the Chang-Hinton formula~\cite{Chang82,Chang86} is shown in Fig.~\ref{fig:JET_qH_fluxes_against_CH}. Including the effect of tungsten on Deuterium transport, by setting up the $\alpha$ parameter with the same profile of charge $\langle Z\rangle$ used in the simulation, recovers the simulation result somewhat but still overestimates the increase of heat transport in the presence of tungsten ions, for this plasma.

\section{Discussion on up-coming modeling of atomic interaction between tungsten bundles}
By atomic interaction we refer to the processes of ionization and recombination, which would be modeled by a transfer of density between bundles of different charge states. For instance, if all the tungsten ions were initially deposited on the bundle $Z=14.1$, we would need a mechanism to transfers this tungsten in  between the different bundles. This approach is going to be needed in a longer time simulation and will be reported in a future publication. 

As can be seen from Fig.~\ref{fig:fluxes_impurities_multiW}(a),  the radial distribution of the bundled tungsten particles has not changed much from the initial distribution in this simulation. Therefore, the radial profile of charge $\langle Z\rangle$ does not change much and would not require the inclusion of atomic interactions between tungsten bundles to consistently maintain this radial profile.

On the other hand, the rapid formation of  large poloidal asymmetries can affect the average charge state $\langle Z \rangle$ poloidally if the phase of these asymmetries depends on the charge states of the bundle. In our current simulation we have a dominant bundle at each location and it is so dominant that even if different asymmetries arise, the average $\langle Z\rangle$ remains relatively stable. The only potential perturbation is where bundles of neighboring charge have similar density but opposite phase asymmetry which is not observed in this work.

Finally, the modeling of atomic interaction will be necessary to study the sputtering of tungsten from the wall and its penetration across the separatrix towards the core. Indeed, a bundle of low charge state will be saturated in tungsten ions in the SOL and will need to feed a bundle of neighboring charge by ionization. The modeling of such atomic interaction based on ADAS database between bundles have been initiated and will be reported in the near future.

\section{Conclusion}
\label{sec:conclusion} 
A basic study of the collisional transport of  tungsten ions through the entire volume of JET geometry has been conducted using a model H-mode plasma with the magnetic drift direction downward towards the X-point. {This model plasma is inspired from a carbon-wall plasma in which we added some tungsten.} The radial profile of the average charge $\langle Z\rangle$ of the tungsten ions is modeled by using four bundles of different electric charge. The density and electric charge of each bundle is initialized from a Coronal approximation. Tungsten ions have higher-Z in the core ($\simeq45$) and lower-Z in the edge ($\simeq10$).

The radial transport of tungsten ions in the core region is found to be significantly different from the one in the edge. The high-Z tungsten particles move out from the core toward the pedestal top area, whereas the low-Z tungsten particles move inward from the separatrix toward the pedestal top and core. This organization of the fluxes of tungsten ions, which leads to an accumulation of tungsten in the pedestal top, is found to be significantly influenced by the temperature screening effect and the poloidal asymmetries. High-Z tungsten ions behave as if they are more screened than the low-Z tungsten ions. From the operation point of view, this indicates that increasing the electron temperature will increase the tungsten ionization state and thus increase the outgoing flux of tungsten ions from the core.

The up-down asymmetries of large amplitude observed on the density of tungsten ions are found to be correlated with the direction of the tungsten particle flux. Indeed, in these neoclassical simulations where the particle flux is dominated by the grad-B drift contribution and where the grad-B drift is pointing downward, an accumulation of tungsten at the top leads to an inward flux of particles and an accumulation of tungsten at the bottom leads to an outward flux of particles. This is observed for all tungsten ions in the high collisional regime. Only the low-Z tungsten ions that have a lower collisionality (plateau) in the core do not follow this trend. 

{The reversal of the up-down asymmetries and associated parallel friction force in the pedestal appears to be caused by the modulation of the deuterium and tungsten parallel flows near the separatrix, where the ion-orbit loss effects drive this parallel flow dynamics. The possible mitigation of tungsten penetration via the interaction of strong rotation and ion-orbit loss effects will be studied in the near future. The influence of poloidal asymmetry caused by neutral beam heating\cite{Chang83} or via ICRH\cite{Choe95,Angioni14} could also be considered.

As discussed, a simple analysis of the influence of asymmetries on the temperature screening effect showed a correlation between the asymmetry factor $\frac{\mathcal{C}_U}{\mathcal{C}_G}$ and the direction of the particle flux. This preliminary analysis suggests that the modification of the parallel flow via ion-orbit losses modulates the up-down asymmetry and in turn reduces the temperature screening effect. This reduction of the temperature screening effect would explain the strong penetration of tungsten from the SOL to the pedestal top.

Finally, modeling the global organization of the tungsten ions transport requires a global modeling of the radial profile of tungsten charge $\langle Z\rangle$ and cannot be captured by a single tungsten species with a fixed charge $Z$. As demonstrated, this charge profile affects the global organization of asymmetries and radial transport. This point is well emphasized by low-Z tungsten ions that have an opposite direction of asymmetries in the core compared to the most abundant high-Z ions at this location. A single gyrokinetic species with fixed charge $Z$ enables a good estimate of the local transport only at locations where this charge $Z$ is close to the most abundant one, \textit{i.e.}, when $Z\simeq\langle Z\rangle$. Time-evolution of the modeled charge $\langle Z\rangle$ in our short-time simulation is weak. The radial neoclassical transport is not strong enough over our simulation time scale to change the density profiles of the tungsten bundles. The inclusion of atomic interactions between bundles will be considered in a future work.}

\section{Acknowledgement}
The authors would like to thanks Per Helander, Patrick Maget, and Pierre Manas for fruitful discussions, Aaron Scheinberg for his help with the XGC code, and Norbert Podhorszki for his help with the ADIOS library.

This research was supported by the SciDAC project “High-fidelity Boundary Plasma Simulation,” and the Exascale Computing Project (No. 17-SC-20-SC), a collaborative effort of the U.S. Department of Energy Office of Science and the National Nuclear Security Administration.

This research used {\color{black}INCITE} resources of the Oak Ridge Leadership Computing {\color{black}Facilities} at the Oak Ridge National Laboratory {\color{black}(OLCF) and Argonne National Laboratory (ALCF),}  and resources of the National Energy Research Scientific Computing Center (NERSC), which are supported by the Office of Science of the U.S. Department of Energy under Contract Nos. DE-AC05-00OR22725 and DE-AC02-05CH11231, respectively.

This work was supported by the U.S. Department of Energy under contract number DE-AC02-09CH11466. The United States Government retains a non-exclusive, paid-up, irrevocable, world-wide license to publish or reproduce the published form of this manuscript, or allow others to do so, for United States Government purposes. 

This work has been carried out within the framework of the EUROfusion Consortium, funded by the European Union via the Euratom Research and Training Program (Grant Agreement No 101052200 — EUROfusion). Views and opinions expressed are however those of the author(s) only and do not necessarily reflect those of the European Union or the European Commission. Neither the European Union nor the European Commission can be held responsible for them.

\bibliographystyle{apsrev4-1}
\bibliography{bibliography}
\end{document}